\documentclass[a4paper]{article}

\usepackage[english]{babel}
\usepackage[utf8]{inputenc}
\usepackage{amsmath}
\usepackage{graphicx}
\usepackage[colorinlistoftodos]{todonotes}
\usepackage{xcolor}
\usepackage{multirow}

\title{Text-Independent Speaker Identification Using Audio Looping With Margin Based Loss Functions}

\author{ 
   \large Elliot Q. C. Garcia \\[-3pt]
   \normalsize Universidade Federal Rural de Pernambuco  \\[-3pt]
    \normalsize  elliot.qcgarcia@ufrpe.br\\[-3pt]
   \normalsize Futuro Tech  \\[-3pt]
    \normalsize  elliot.garcia@futuroai.tech\\[-3pt]
  \and
   \large Nicéias Silva Vilela \\[-3pt]
   \normalsize Universidade Federal Rural de Pernambuco  \\[-3pt]
    \normalsize	niceias.vilela@ufrpe.br \\[-3pt]
\and
   \large  Kátia Pires Nascimento do Sacramento \\[-3pt]
   \normalsize Universidade Regional do Cariri\\[-3pt]
    \normalsize	Katia.pires@urca.br \\[-3pt]
\and
   \large Tiago A. E. Ferreira \\[-3pt]
   \normalsize Universidade Federal Rural de Pernambuco \\[-3pt]
    \normalsize	tiago.espinola@ufrpe.br \\[-3pt]
}

\date{\today}

\begin{document}
\maketitle

\begin{abstract}

Speaker identification has become a crucial component in various applications, including security systems, virtual assistants, and personalized user experiences. In this paper, we investigate the effectiveness of CosFace Loss and ArcFace Loss for text-independent speaker identification using a Convolutional Neural Network architecture based on the VGG16 model, modified to accommodate mel spectrogram inputs of variable sizes generated from the Voxceleb1 dataset. Our approach involves implementing both loss functions to analyze their effects on model accuracy and robustness, where the Softmax loss function was employed as a comparative baseline. Additionally, we examine how the sizes of mel spectrograms and their varying time lengths influence model performance. The experimental results demonstrate superior identification accuracy compared to traditional Softmax loss methods. Furthermore, we discuss the implications of these findings for future research.

\end{abstract}

\section{Introduction} \label{Introduction}

Speaker Recognition (SR) has become a widely used application for security systems, virtual assistants, and personalized user experiences~\cite{9519395, magalhaes2021voice}. Traditional methods, such as Mel-Frequency Cepstral Coefficients (MFCCs)~\cite{kamruzzaman2010speaker, sinith2010novel}, have been widely used in SR but often struggle with noise sensitivity and limited discriminative power. With the advent of deep learning, spectrograms have emerged as an alternative for speaker identification~\cite{yadav2018learning, anand2019shotspeakerrecognitionusing, an2019deep}. Spectrograms provide a time-frequency representation of speech signals, capturing both temporal and spectral information. These spectrograms are often combined with Convolutional Neural Networks (CNNs) where the CNNs are trained in a supervised process guided by classification loss functions. Current prevailing classification loss functions for SR systems are mostly based on the Softmax Loss function. However, Softmax Loss does not explicitly optimize feature embedding to enforce higher similarity for intra-class samples and diversity for inter-class samples. To remedy this, we decided to approach this problem using two different loss functions and comparing them to the traditional Softmax approach, namely Arcface and CosFace Loss. 

SR systems can be modeled either as Text-Dependent or Text-Independent systems. Text-Dependent SR relies on the user uttering whatever is being prompted, while text-independent SR provides a more flexible approach where there is no constraint on the user of what can be said, and as such, has broader applications.

SR encompasses voice-based technologies designed to identify or verify individuals using their vocal characteristics. Although it can be divided into several subcategories, it is generally divided into three primary categories:
\begin{itemize}
    \item Speaker Verification (SV). Verifies whether a speaker's claimed identity is true. The system compares the speaker's voice to a stored voiceprint of the claimed identity to confirm their identity~\cite{juang2003digital}. Speaker verification is commonly used in authentication systems, such as secure access to privileged information, devices, or accounts;

    \item Speaker Identification (SI). Determines the identity of an unknown speaker from a group of known speakers. The system compares the input voice to a database of voiceprints and identifies the closest match. SI can then also be split into two approaches, the closed-set approach and the open-set approach. In the case of closed-set identification, the speakers are initially already enrolled in the database, and the system assumes that the current unknown speaker is already enrolled in the database. In the case of open-set identification, the speaker is not always enrolled in the database, as such the system needs to be capable of rejecting a speaker~\cite{campbell2002speaker}. SI is widely used in applications such as law enforcement, voice assistants, and call center analytics~\cite{furui2010speaker};

    \item Speaker Classification (SC). Distinguishes and classifies speakers based on specific characteristics such as age, gender and health. It is commonly used in scenarios such as demographic analysis and targeted marketing\cite{qawaqneh2017deep}.

\end{itemize}

Each category faces unique challenges and applications. For instance, speaker verification must handle variability in a person's voice due to emotional states or background noise, while SI requires robust matching algorithms to distinguish between similar voices, often necessitating large and diverse databases to improve accuracy. SC, on the other hand, requires sophisticated feature extraction techniques and machine learning models to accurately capture and analyze subtle vocal traits.

The underlying system for SI works similarly to a fingerprint matching process. Looking at the spectral content, the system can analyze the unique features of a person's voice and match it against a database. These features are highly distinctive, capturing the physiological and behavioral characteristics of an individual's voice~\cite{morandieffect}.


The present paper makes three main contributions. First, we evaluate the comparative effectiveness of the advanced loss functions ArcFace and CosFace against the traditional Softmax loss for SI using modified VGG16 architectures with mel-spectrogram inputs. It is worth noting that the time duration of the voice sample can influence the mel-spectrograms' capacity to represent the speaker. Therefore, we also investigate two different sample durations: 3 and 10 seconds. To extend the sample duration, we propose a new method based on time-loop repetition, where the original sample is repeated until the desired duration is reached. Accordingly, the second contribution focuses on the length of the voice samples. We demonstrate that audio looping techniques significantly improve identification accuracy by extending shorter recordings to optimal lengths for feature extraction. Finally, we systematically analyze how varying mel-spectrogram image dimensions ($224\times 224\times 3$, $448\times 448\times 3$, and $432\times 288\times 3$) affect model performance, identifying optimal input configurations that balance computational efficiency with identification accuracy.


The paper is organized as follows. Section \ref{Related Work} presents related works on SI systems utilizing CNNs with spectrogram inputs and establishes the context for our contributions. Section \ref{Methodology} describes our methodology, including preprocessing techniques applied to the VoxCeleb1 dataset\footnote{https://www.robots.ox.ac.uk/\text{˜}vgg/data/voxceleb/vox1.html}~\cite{nagrani2017voxceleb}, the advanced loss functions employed and their theoretical advantages, and the architectural modifications made to the VGG16 model. Section \ref{Results and Discussion} presents experimental results and discusses the impact of audio length and looping techniques, the comparative effectiveness of different loss functions, and the effects of varying mel-spectrogram dimensions on identification accuracy. Section \ref{Conclusion} concludes the paper with a summary of findings and implications for future research.

\section{Related Work} \label{Related Work}

This section is to briefly review several related or similar works in line with this paper. The technique of generating speech-based spectrograms is implemented during the feature extraction stage, while CNN are used during the classification stage.

Nagrani \textit{et al.}~\cite{nagrani2017voxceleb} originally introduced and tested the VoxCeleb1 dataset. They randomly selected three second segments from each speech and generated mel-spectrogram images of size $512\times 300\times 3$. For classification, they used a modified VGG CNN, having changed the final maxpool layer to avgpool. They achieved a top-1 accuracy of $80.5\%$ with the traditional Softmax loss function.

Anand \textit{et al.}~\cite{anand2019shotspeakerrecognitionusing} propose a $3$-second random selection of each speech and then convert them to spectrogram images of size $128\times 300\times 1$. In the classification stage, the spectrograms are classified by different CNNs, such as VGG, ResNet, and CapsuleNet. They employed a combination of Margin Loss for training the capsule networks and Prototypical Loss for generalizing under unseen speakers. The ResNet classifier for $200$ classes from the Voxceleb1 dataset has the best result of top-1 = $71.8\%$.

An \textit{et al.}~\cite{an2019deep} improved upon the original SI system's accuracy \cite{nagrani2017voxceleb} by altering the VGG network, specifically replacing the final max pooling layer with an average pooling layer and adding a self-attention layer before this pooling layer. This modification enabled the system to manage variable-length segments effectively. The study employed a combination of a penalization term alongside the traditional cross-entropy loss. For feature extraction, the researchers utilized mel-spectrograms, extracted from three-second audio segments, with dimensions of $512\times 300 \times 3$. The resulting accuracy on the VoxCeleb1 dataset was a top-1 of $= 90.8\%$.

Similarly, Yadav and Rai~\cite{yadav2018learning} extracted mel-spectrograms from three-second utterances. However, they were generated with size $301 \times 161 \times 1$. One of the methods used in the classification stage was a VGG13-based CNN by adding a batch normalization layer after every convolutional layer. The system employs joint supervision of Softmax and Center Loss, achieving a top-1 accuracy of $89.5\%$  on the VoxCeleb1 dataset.

Sharif \textit{et al.}~\cite{sharif2025improving} initially extracted mel-spectrogram images of size $535\times 678\times 3$, resizing them afterwards to $100\times 128\times 3$. The study optimized the VGG-13 architecture for speaker recognition by reducing the number of convolutional layers from 10 to 5 and changing the pooling layers from max pooling to average pooling. They also added batch normalization layers after each average pooling layer and decreased the dropout layers from 10 to just 1. Furthermore, the fully connected layers were modified from three layers to a single layer with $1251$ hidden neurons, matching the number of speakers in the VoxCeleb1 dataset. The study primarily utilized triple loss, n-pair loss, and angular loss, achieving a top-1 accuracy of $91.17\%$.


The works discussed in this section demonstrate several trends. Researchers have consistently modified VGG network architectures, achieving positive results in SI tasks. They have employed advanced loss functions, resulting in improved model generalization capabilities. The reported accuracy ranges from $71.8\%$ to $91.17\%$ across different studies \cite{nagrani2017voxceleb, anand2019shotspeakerrecognitionusing, an2019deep, yadav2018learning, nugroho2022enhanced}. All studies utilized VoxCeleb1 as their benchmark dataset while implementing various data augmentation techniques and experimenting with different spectrogram dimensions.
Building upon these established approaches, our work contributes to this research landscape by implementing modifications to the VGG16 architecture, systematically comparing two advanced loss functions, namely ArcFace and CosFace, and experimenting with multiple mel-spectrogram dimensions. Additionally, we introduce an audio looping technique as a data augmentation method for the VoxCeleb1 dataset, achieving a competitive top-1 accuracy of $83.15\%$ that falls within the established performance range while providing new insights into optimal input configurations.

\section{Methodology} \label{Methodology}

\subsection{Dataset} \label{Dataset}

For this study, we used the VoxCeleb1 dataset~\cite{nagrani2017voxceleb}, a widely used benchmark for SR and verification tasks. VoxCeleb1 contains over $100,000$ utterances from $1,251$ speakers, collected from publicly available interview videos on platforms such as YouTube~\cite{nagrani2017voxceleb}.

The audio samples in VoxCeleb1 are recorded at a sample rate of $16$kHz, which is standard for speech processing tasks. The dataset includes recordings in multiple languages, with speakers of diverse accents, ages, and genders, being made up of 688 males and 563 females. 

The utterances are recorded in various environments, from outdoor stadiums to quiet indoor studios, with almost all of them having some form of background noise, such as laughter, background chatter, and overlapping speaking.

For our experiment, we used a train-validation-test split of $70\%$, $15\%$, $15\%$ as illustrated in Table~\ref{voxcelebsplit}.

\begin{table}
\caption{Information about the database used for experimentation.}
\centering
{
\begin{tabular}{l|l}
\hline
Parameters &  Voxceleb1\\\hline
Total number of speakers &  1251(688 male, 563 female)\\
Total utterances &  153,516\\
Train utterances & 108,018\\
Validation utterances & 23,066\\
Test utterances & 22,432\\
\hline
\end{tabular}
}
\label{voxcelebsplit}
\end{table}

\subsection{Preprocessing} \label{Preprocessing}

\begin{figure}[!h]
\centering
\includegraphics[width=.8\linewidth,height=.13\textheight]{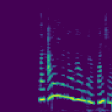}
\caption{Spectrogram image of audio before silence removal.} \label{Before removal}
\end{figure}

A lot of the audios start or end with silence. Figure \ref{Before removal} shows an example where there is a silent situation at the beginning (region with only purple color). We start by reducing this, ensuring that the system does not waste resources by processing empty data.

\subsubsection{Data Augmentation} \label{Data Augmentation}

\begin{figure}
\centering
\includegraphics[width=.8\linewidth,height=.25\textheight,keepaspectratio]{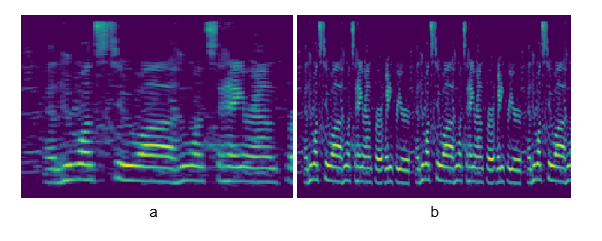}
\caption{Visual representation of mel-spectrograms (a) before and (b) after looping. The looping process extends shorter audio segments to meet the required duration.} \label{compmelspec310}
\end{figure}

After reducing the silence from the audio signals, many of the recordings did not reach the desired duration. To address this, we implemented an audio looping technique to extend these recordings to the required length, as illustrated in Figure~\ref{compmelspec310}. 

This enabled the creation of two distinct dataset versions: (a) consisting of three-second audio clips and (b) with ten-second clips. Our experimental setup utilized these target durations, with three seconds chosen as it is a common standard in SI research, and ten seconds selected based on observations of diminishing returns where extending duration further yielded no noticeable accuracy improvements.

It is important to note the specific implementation of our audio looping method: looping was applied only at the end of an audio segment if it was shorter than the target duration. If an audio sample already met or exceeded the target length (e.g., 10 seconds), no looping was applied, and for longer samples (e.g., 20 seconds), a 10-second segment was extracted. This approach means that the effectiveness of the audio looping method in more consistent environments or with different looping strategies requires further investigation.

\subsubsection{Mel-spectrogram} \label{Mel-spectrogram}

\begin{figure}[!h]
\centering
\includegraphics[width=.8\linewidth,height=.25\textheight,keepaspectratio]{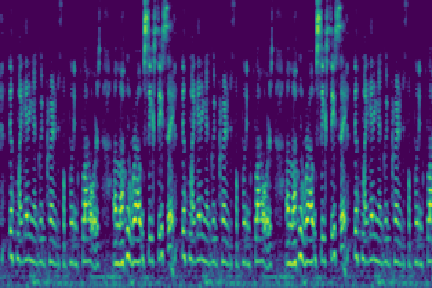}
\caption{A typical Mel-spectrogram image used to represent a speech signal.} \label{melspectrogram}
\end{figure}

Spectrograms are a graphical representation of a signal's frequencies. When applied to an audio or voice signal, it generates a visual representation of signals, depicting the distribution of energy across time and frequency. They are generated using the Short-Term Fourier Transform (STFT)\cite{badshah2019deep}, which converts raw audio signals into a two-dimensional representation, where the x-axis represents time, the y-axis represents frequency, and the color intensity represents the energy amplitude, as can be seen in Figure~\ref{melspectrogram}. 

Spectrograms, particularly mel-spectrograms, are robust to noise and variations in recording conditions, as shown by Nagrani \textit{et al.}~\cite{nagrani2017voxceleb}, Lambamo \textit{et al.}~\cite{lambamo2022analyzing}, and Saritha \textit{et al.}~\cite{saritha2024deep}. In these works, the authors showed that spectrograms can mitigate the effects of background noise by focusing on the frequency patterns that are most relevant to the characteristics of the speaker. The conversion of spectrograms into the Mel scale is done because the Mel scale is designed to mimic the way humans perceive sound, with a non-linear frequency resolution that assigns more weight to lower frequencies. Lower frequency components have been shown to be more distinguishing between speakers in SI systems\cite{lu2008investigation, zouhir2025power}. 

Using the audio data we previously generated, we convert the recordings into ten-second Mel-spectrograms. These are then resized into three specific dimensions: $224\times 224\times 3$, $448\times 448\times 3$, and $432\times 288\times 3$. The same process is applied to the dataset of three-second audio recordings. Additionally, we apply mean and variance normalization on a per-speaker basis to ensure consistent processing across the dataset \cite{nagrani2017voxceleb}.
 
\subsection{Loss functions} \label{Loss functions}

Much research has been conducted using the cross-entropy loss function, commonly known as Softmax loss \cite{nagrani2017voxceleb, an2019deep}. 
The Softmax loss function for a batch of $n$ samples is mathematically defined as:

\begin{equation}
L_{\text{AMS}} = -\frac{1}{n} \sum_{i=1}^{n} \log \left( P(y = y_i | x_i ; W) \right)
\end{equation}
where $\left( P(y = y_i | x_i ; W) \right)$ is the probability of $y$ is the class $y_i$, given the sample $x_i$ and the weight vector $W$.

\begin{figure}[h!]
\centering
\includegraphics[width=.8\linewidth,height=.25\textheight,keepaspectratio]{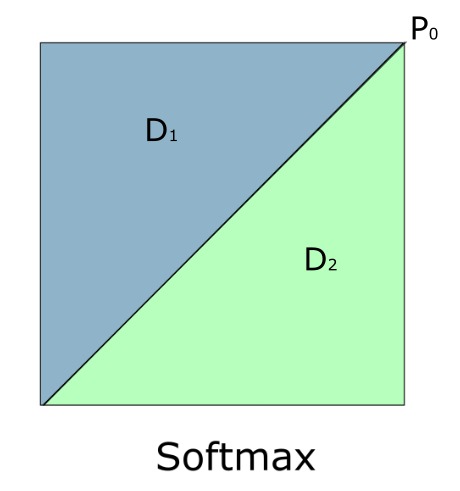}
\caption{Traditional Softmax's decision boundary meets at $P_0$. $D_1$ and $D_2$ represent class 1 and class 2, respectively.} 
\label{Softmaxdecisionboundary}
\end{figure}

While these approaches have yielded positive results, Softmax presents several limitations for SI. A primary drawback is its failure to explicitly enforce discriminative margins between different speaker classes. This can result in learned embeddings for distinct speakers being too close in the feature space, hindering the model's ability to differentiate them, particularly under noisy conditions or with limited training data. Furthermore, Softmax does not directly optimize for intra-class compactness; while it pushes embeddings away from decision boundaries, it doesn't strongly encourage embeddings from the same speaker to cluster tightly together as seen in Figure~\ref{Softmaxdecisionboundary}. This can lead to more spread-out representations for a single speaker, potentially reducing the robustness of the SI system. An additional concern is that Softmax treats all classes equally, which can be problematic in SI, where the number of speakers (classes) can be very large, potentially leading to imbalanced class distributions and a bias towards more frequent speakers in the training data.

To address these limitations, our study investigates the effectiveness of alternative loss functions: CosFace and ArcFace\cite{wang2018cosface, deng2019arcface}. These functions are designed to directly improve class separation and intra-class compactness, which are crucial for robust SI. We focus on enhancing intra-class similarities by employing these functions, which, although originally developed for facial recognition, have demonstrated significant potential for SI tasks. Both CosFace and ArcFace work by enhancing intra-class similarities through the imposition of angular or margin-based constraints \cite{wang2018cosface, wang2018additive, deng2019arcface}.


\subsection{CosFace} \label{Cosface}

CosFace Loss is an enhanced version of the traditional Softmax loss, designed to improve the discriminative power of deep neural networks by imposing a fixed angular margin between classes~\cite{wang2018additive}. This is achieved by modifying the target logit through the addition of a margin to the cosine similarity score of the target class.

The CosFace loss introduces an additive margin to the cosine similarity score of the target class, formulated as:
\begin{equation}
\psi(x) = x - m
\end{equation}
where $x = \cos \theta_{yi}$ is the cosine similarity between the feature vector of the target class, and $m$ is the additive margin. This formulation ensures that the decision boundary is pushed away from the target class by a fixed margin, thereby improving class separation.

\begin{figure}[h!]
\centering
\includegraphics[width=.8\linewidth,height=.25\textheight,keepaspectratio]{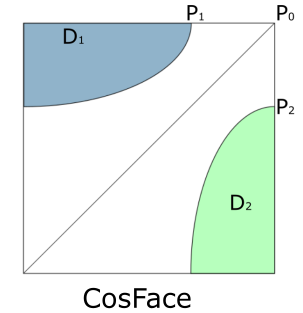}
\caption{A comparison between traditional Softmax's decision boundary and CosFace's decision boundary. Softmax's decision boundary meets at $P_0$ whereas CosFace's decision boundary for class 1 is at $P_1$ and for class 2 it is at $P_2$. $D_1$ and $D_2$ represent class 1 and class 2, respectively.} 
\label{Cosfacedecisionboundary}
\end{figure}

The geometric interpretation of this concept is illustrated in Figure~\ref{Cosfacedecisionboundary}. While traditional Softmax loss places the decision boundary at a single hyperplane  $P_{o}$ (two-class problem), CosFace introduces a marginal region, shifting the boundary away from the target class. This shift is calculated as $m = (W_1 - W_2)^TP_1$. Where $W_1$ and $W_2$ are the weight vectors of the two classes, and $P_1$ is the decision boundary for class $1$. This fixed angular margin leads to more discriminative and compact features. 

The CosFace loss function for sample $x_i$ is mathematically defined as:
\begin{equation}
L_{\text{CF}} = -\log \left( \frac{e^{s(\cos (\theta_{y_{i}}) - m)}}{e^{s(\cos (\theta_{y_{i}}) - m)} + \sum_{j \neq y_i} e^{s \cos \theta_j}} \right)
\end{equation}

where:
\begin{itemize}
    \item $s$ is the scaling factor, which amplifies the separation between classes;
    \item $m$ is the additive margin, which enforces a fixed angular separation;
    \item $\cos ( \theta_{y_{i}} )$ is the cosine similarity between the features and the target weights.
\end{itemize}

For our study, we used $s = 22$ and $m = 0.2$, which were chosen to optimize the trade-off between class separation and model convergence.

\subsection{ArcFace} \label{ArcFace}

The ArcFace loss function, introduced by Deng \textit{et al.}~\cite{deng2019arcface}, was originally used for face verification, but has seen some recent use in speaker recognition technology \cite{guo2021speaker}. This function is specifically designed to optimize the geodesic distance margin on a hypersphere by introducing an additive angular margin that pushes the decision boundary further from the target class, thereby enhancing class separation.

\begin{figure}[h!]
\centering
\includegraphics[width=.8\linewidth,height=.25\textheight,keepaspectratio]{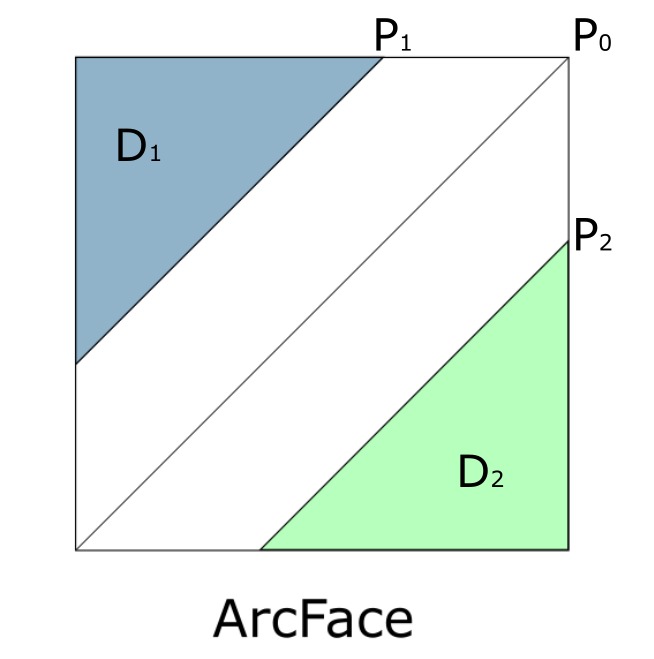}
\caption{A comparison between traditional Softmax's decision boundary and ArcFace's decision boundary. Softmax's decision boundary meets at $P_0$, whereas ArcFace's decision boundary for class 1 is at $P_1$ and for class 2 it is at $P_2$. $D_1$ and $D_2$ represent class 1 and class 2, respectively.} 
\label{arcfacedecisionboundary}
\end{figure}

ArcFace optimizes speaker embeddings through a dual approach: maximizing intra-class compactness while simultaneously increasing inter-class separation as seen in Figure~\ref{arcfacedecisionboundary}. This optimization strategy has seen some success in improving speaker distinction, particularly when paired with large-scale datasets where subtle differences between speakers must be accurately captured.

The integration of ArcFace with SincNet architectures \cite{guo2021speaker} has enabled direct feature extraction from raw speech signals. When combined with dual attention mechanisms, this approach delivers performance improvements, especially in short-utterance scenarios where limited audio data is available for SI.

ArcFace demonstrates the capability to effectively handle difficult operational conditions, including short utterances and cross-domain mismatches. In far-field speaker verification applications, the loss function has been successfully adapted to address domain mismatches \cite{lin2022cross}, with optimized margin penalties during training significantly enhancing overall system performance.

The practical advantages of ArcFace extend to various challenging environments. Models incorporating ArcFace have shown reduced error rates in short speech scenarios, making them effective for real-world applications where audio samples are typically brief. Additionally, the enhanced feature extraction and classification capabilities contribute to superior performance in noisy environments, addressing one of the most common challenges in SI systems.

Comparative studies have demonstrated that ArcFace-based models consistently achieve lower error rates than traditional methods \cite{tang2022bimodal}, underscoring their effectiveness in SI tasks. Furthermore, the loss function's inherent ability to stabilize training processes and enhance feature learning establishes it as an invaluable tool in deep learning-based speaker recognition systems \cite{tang2022bimodal}. 

For a sample ($x_i,y_i$), where $x_i$ is the input sample and $y_i$ is the target label, the mathematical formulation of the ArcFace loss function is as follows:

\begin{equation}
L_{\text{AF}}(x_i) = -log \left( \frac{e^{s(\cos(\theta_i + m))}}{e^{s(\cos(\theta_i + m))} + \sum_{j\neq y_i} e^{s(\cos(\theta_j))}}\right)
\end{equation}

where:
\begin{itemize}
    \item $s$ is the scaling factor;
    \item $m$ is the additive angular margin;
    \item $\cos (\theta_{y_{i}})$ is the angle between the feature vector and the target weights.
\end{itemize}

In our implementation, the initial scaling factor $s$ was set to $s = 22$. The additive angular margin $m$ was set to $0.2$. This configuration was chosen to ensure more compact and discriminative features.

\subsection{Recognition Model} \label{Model}

\subsubsection{VGG16} \label{VGG16}

\begin{table}[]
\caption{Original VGG16 Architecture. This table details the sequential layers, kernel sizes, strides, and activation functions characteristic of the VGG16 network as described by Simonyan and Zisserman\cite{simonyan2014vgg}.}{
\resizebox{\columnwidth}{!}{
\begin{tabular}{|cccccc|}
\hline
\multicolumn{6}{|c|}{Original VGG16}                                                                                                                                                                                        \\ \hline
\multicolumn{1}{|c|}{Layer}                                         & \multicolumn{1}{c|}{Feature Map} & \multicolumn{1}{|c|}{Size}            & \multicolumn{1}{c|}{Kernel Size} & \multicolumn{1}{c|}{Stride} & Activation \\ \hline
 \multicolumn{1}{|c|}{Image}           & \multicolumn{1}{c|}{1}           & \multicolumn{1}{c|}{224 x 224 x3}    & \multicolumn{1}{c|}{-}           & \multicolumn{1}{c|}{-}      & -          \\ \hline
 \multicolumn{1}{|c|}{2 X Convolution} & \multicolumn{1}{c|}{64}          & \multicolumn{1}{c|}{224 x 224 x 64}  & \multicolumn{1}{c|}{3x3}         & \multicolumn{1}{c|}{1}      & relu       \\ \hline
 \multicolumn{1}{|c|}{Max Pooling}     & \multicolumn{1}{c|}{64}          & \multicolumn{1}{c|}{112 x 112 x 64}  & \multicolumn{1}{c|}{3x3}         & \multicolumn{1}{c|}{2}      & relu       \\ \hline
 \multicolumn{1}{|c|}{2 X Convolution} & \multicolumn{1}{c|}{128}         & \multicolumn{1}{c|}{112 x 112 x 128} & \multicolumn{1}{c|}{3x3}         & \multicolumn{1}{c|}{1}      & relu       \\ \hline
 \multicolumn{1}{|c|}{Max Pooling}     & \multicolumn{1}{c|}{128}         & \multicolumn{1}{c|}{56 x 56 x 128}   & \multicolumn{1}{c|}{3x3}         & \multicolumn{1}{c|}{2}      & relu       \\ \hline
 \multicolumn{1}{|c|}{2 X Convolution} & \multicolumn{1}{c|}{256}         & \multicolumn{1}{c|}{56 x 56 x 256}   & \multicolumn{1}{c|}{3x3}         & \multicolumn{1}{c|}{1}      & relu       \\ \hline
 \multicolumn{1}{|c|}{Max Pooling}     & \multicolumn{1}{c|}{256}         & \multicolumn{1}{c|}{28 x 28 x 256}   & \multicolumn{1}{c|}{3x3}         & \multicolumn{1}{c|}{2}      & relu       \\ \hline
 \multicolumn{1}{|c|}{3 X Convolution} & \multicolumn{1}{c|}{512}         & \multicolumn{1}{c|}{28 x 28 x 512}   & \multicolumn{1}{c|}{3x3}         & \multicolumn{1}{c|}{1}      & relu       \\ \hline
 \multicolumn{1}{|c|}{Max Pooling}     & \multicolumn{1}{c|}{512}         & \multicolumn{1}{c|}{14 x 14 x 512}   & \multicolumn{1}{c|}{3x3}         & \multicolumn{1}{c|}{2}      & relu       \\ \hline
 \multicolumn{1}{|c|}{3 X Convolution} & \multicolumn{1}{c|}{512}         & \multicolumn{1}{c|}{14 x 14 x 512}   & \multicolumn{1}{c|}{3x3}         & \multicolumn{1}{c|}{1}      & relu       \\ \hline
 \multicolumn{1}{|c|}{Max Pooling}     & \multicolumn{1}{c|}{512}         & \multicolumn{1}{c|}{7 x 7 x 512}     & \multicolumn{1}{c|}{3x3}         & \multicolumn{1}{c|}{2}      & relu       \\ \hline
 \multicolumn{1}{|c|}{Dense}           & \multicolumn{1}{c|}{-}           & \multicolumn{1}{c|}{25088}           & \multicolumn{1}{c|}{-}           & \multicolumn{1}{c|}{-}      & relu       \\ \hline
 \multicolumn{1}{|c|}{Dense}           & \multicolumn{1}{c|}{-}           & \multicolumn{1}{c|}{4096}            & \multicolumn{1}{c|}{-}           & \multicolumn{1}{c|}{-}      & relu       \\ \hline
 \multicolumn{1}{|c|}{Dense}           & \multicolumn{1}{c|}{-}           & \multicolumn{1}{c|}{4096}            & \multicolumn{1}{c|}{-}           & \multicolumn{1}{c|}{-}      & relu       \\ \hline
\multicolumn{1}{|c|}{Dense}           & \multicolumn{1}{c|}{-}           & \multicolumn{1}{c|}{1000}            & \multicolumn{1}{c|}{-}           & \multicolumn{1}{c|}{-}      & Softmax    \\ \hline
\end{tabular}}
}
\label{tableVGG16}
\end{table}

The VGG16 model is a CNN comprising 16 weight layers: 13 convolutional layers and 3 fully connected layers. These layers are organized into five blocks, each followed by a max-pooling layer to progressively reduce the spatial dimensions of the feature maps~\cite{simonyan2014vgg} as can be seen in Table \ref{tableVGG16}. The core of the architecture can be broken down as follows:

\begin{itemize}
    \item Convolutional Layers: The first two blocks contain two convolutional layers each, while the last three blocks contain three convolutional layers each. Each convolutional layer applies a $3 \times 3$ filter with stride $1$ and padding to preserve spatial resolution, followed by a ReLU activation function;
    \item Pooling Layers: A max-pooling layer with a $2\times 2$ filter and stride $2$ is applied after each block to downsample the feature maps.
\end{itemize}

\subsubsection{Modified VGG16} \label{Modified VGG16}

\begin{table}[]
\caption{Proposed Modified VGG16 Architecture. 
It highlights the proposed modifications to the VGG16 convolutional base, retaining its core feature extraction blocks. Notable adaptations include the integration of global average pooling, a custom classifier yielding 256-dimensional embeddings, and a final classification head for SI. The embeddings are further normalized using L2 normalization for cosine similarity.}{
\resizebox{\columnwidth}{!}{%
\begin{tabular}{|cccccc|}
\hline
\multicolumn{6}{|c|}{Modified VGG16}                                                                                                                                                                                                                   \\ \hline
Layer                                               & \multicolumn{1}{c|}{Feature Map} & \multicolumn{1}{c|}{Size}                         & \multicolumn{1}{|c|}{Kernel Size} & \multicolumn{1}{c|}{Stride} & Activation         \\ \hline
\multicolumn{1}{|c|}{Image}                 & \multicolumn{1}{c|}{1}           & \multicolumn{1}{c|}{Any Height x Any   Width x 3} & \multicolumn{1}{c|}{-}           & \multicolumn{1}{c|}{-}      & -                  \\ \hline
 \multicolumn{1}{|c|}{2 X Convolution}       & \multicolumn{1}{c|}{64}          & \multicolumn{1}{c|}{224 x 224 x 64}               & \multicolumn{1}{c|}{3x3}         & \multicolumn{1}{c|}{1}      & relu               \\ \hline
 \multicolumn{1}{|c|}{Max Pooling}           & \multicolumn{1}{c|}{64}          & \multicolumn{1}{c|}{112 x 112 x 64}               & \multicolumn{1}{c|}{3x3}         & \multicolumn{1}{c|}{2}      & relu               \\ \hline
 \multicolumn{1}{|c|}{2 X Convolution}       & \multicolumn{1}{c|}{128}         & \multicolumn{1}{c|}{112 x 112 x 128}              & \multicolumn{1}{c|}{3x3}         & \multicolumn{1}{c|}{1}      & relu               \\ \hline
 \multicolumn{1}{|c|}{Max Pooling}           & \multicolumn{1}{c|}{128}         & \multicolumn{1}{c|}{56 x 56 x 128}                & \multicolumn{1}{c|}{3x3}         & \multicolumn{1}{c|}{2}      & relu               \\ \hline
 \multicolumn{1}{|c|}{2 X Convolution}       & \multicolumn{1}{c|}{256}         & \multicolumn{1}{c|}{56 x 56 x 256}                & \multicolumn{1}{c|}{3x3}         & \multicolumn{1}{c|}{1}      & relu               \\ \hline
 \multicolumn{1}{|c|}{Max Pooling}           & \multicolumn{1}{c|}{256}         & \multicolumn{1}{c|}{28 x 28 x 256}                & \multicolumn{1}{c|}{3x3}         & \multicolumn{1}{c|}{2}      & relu               \\ \hline
 \multicolumn{1}{|c|}{3 X Convolution}       & \multicolumn{1}{c|}{512}         & \multicolumn{1}{c|}{28 x 28 x 512}                & \multicolumn{1}{c|}{3x3}         & \multicolumn{1}{c|}{1}      & relu               \\ \hline
 \multicolumn{1}{|c|}{Max Pooling}           & \multicolumn{1}{c|}{512}         & \multicolumn{1}{c|}{14 x 14 x 512}                & \multicolumn{1}{c|}{3x3}         & \multicolumn{1}{c|}{2}      & relu               \\ \hline
 \multicolumn{1}{|c|}{3 X Convolution}       & \multicolumn{1}{c|}{512}         & \multicolumn{1}{c|}{14 x 14 x 512}                & \multicolumn{1}{c|}{3x3}         & \multicolumn{1}{c|}{1}      & relu               \\ \hline
 \multicolumn{1}{|c|}{Global Avg Pooling 2D} & \multicolumn{1}{c|}{-}           & \multicolumn{1}{c|}{512}                          & \multicolumn{1}{c|}{N/A}         & \multicolumn{1}{c|}{N/A}    & -                  \\ \hline
 \multicolumn{1}{|c|}{Dense}                 & \multicolumn{1}{c|}{-}           & \multicolumn{1}{c|}{1024}                         & \multicolumn{1}{c|}{-}           & \multicolumn{1}{c|}{-}      & relu               \\ \hline
 \multicolumn{1}{|c|}{Dropout (0.3)}         & \multicolumn{1}{c|}{-}           & \multicolumn{1}{c|}{1024}                         & \multicolumn{1}{c|}{-}           & \multicolumn{1}{c|}{-}      & -                  \\ \hline
 \multicolumn{1}{|c|}{Dense}                 & \multicolumn{1}{c|}{-}           & \multicolumn{1}{c|}{256, followed by L2 normalization}                          & \multicolumn{1}{c|}{-}           & \multicolumn{1}{c|}{-}      & relu               \\ \hline
 \multicolumn{1}{|c|}{Dense}                 & \multicolumn{1}{c|}{-}           & \multicolumn{1}{c|}{1251 (num speakers)}          & \multicolumn{1}{c|}{-}           & \multicolumn{1}{c|}{-}      & Softmax \\ \hline
\end{tabular}}
}
\label{VGG16mod}
\end{table}

To enhance the model's performance and adaptability for SI, we proposed the following modifications using PyTorch, as seen in Table \ref{VGG16mod}:

\begin{enumerate}
    \item First, we replaced the fixed size average pooling layer with a global average pooling layer to handle inputs of varying spatial dimensions.
    \item The fully connected classifier was replaced with a custom classifier designed to output embeddings of a specified dimension. The output of the convolutional layers is first flattened and passed through a fully connected layer with $1024$ units, followed by a ReLU activation and dropout of $0.3$ for regularization. Finally, it outputs the optimal 256-dimensional embeddings through another fully connected layer \cite{hajibabaei2018unified};
    \item We added a classification head. This head is a fully connected layer that maps the embeddings to the number of classes, in our case, $1251$;
    \item To ensure the embeddings are normalized to unit vectors, the model is wrapped in a custom class. During the forward pass, the embeddings are normalized using L2 normalization $(||x||^2=1)$, because we use cosine similarity to compare embeddings.
\end{enumerate}

The model parameters were optimized using Stochastic Gradient Descent (SGD) with an initial learning rate of $0.001$, momentum of $0.9$, and weight decay of $1\times 10^{-4}$. To adapt the learning rate during training, we employed a ReduceLROnPlateau scheduler that monitored validation loss. The scheduler reduced the learning rate by a factor of $0.1$ when no improvement in validation loss was observed for $5$ consecutive epochs, with a minimum improvement threshold of $1\times 10^{-4}$ to qualify as meaningful progress. We also incorporated an early stopping mechanism with a patience of $15$ epochs of no improvement, while also enforcing a minimum of $30$ epochs to ensure that the model had sufficient opportunity to converge, even if progress was initially slow.

\section{Experimental results and discussion} \label{Results and Discussion}

\subsection{Impact of audio length and looping technique} \label{Impact of AL and Loop}

The experimental results presented in Table \ref{resultstab} indicate that ten-second audio clips, which utilized the looping technique, consistently achieved higher top-1 accuracy compared to three-second clips, with improvements ranging from $8.65\%$ to $19.64\%$ depending on the specific configuration. For example, ten-second clips achieved up to a top-1 accuracy of $80.79\%$ when using the ArcFace loss function and $83.15\%$ with the CosFace loss function, the traditional Softmax achieved the lowest results of the three, obtaining a top-1 accuracy of $76.41\%$. In contrast, the three-second clips achieved lower accuracies, ranging from $66.28\%$ to $69.82\%$. This significant difference underscores the impact of audio length in capturing speaker-specific features effectively.

The superior performance aligns with previous studies that emphasize the importance of sufficient audio duration for accurate SI~\cite{nagrani2017voxceleb, hajibabaei2018unified}.
However, in this case, the duration of the clips can be attributed to the looping technique, which allowed the model to process a more extended segment of the audio signal.

\begin{table}
\caption{Results of speaker identification accuracy for different loss functions, mel-spectrogram dimensions, and audio lengths. The best accuracy result is highlighted in boldface.}{
\begin{tabular}{|c|l|l|c|}
\hline
\multirow{2}{*}{Loss function} & Mel-Spectrogram  & \multirow{2}{*}{Audio length} & \multirow{2}{*}{Top-1 accuracy}\\
&  Image Dimensions & & \\\hline
\multirow{6}{*}{Softmax} & \multirow{2}{*}{$224\times 224\times 3$} & 3 seconds & $65.76\%$\\\cline{3-4}
 &  & 10 seconds & $75.41\%$\\\cline{2-4}
 & \multirow{2}{*}{$432\times 288\times 3$} & 3 seconds & $67.58\%$\\\cline{3-4}
 &  & 10 seconds & \pmb{$76.41\%$}\\\cline{2-4}
 & \multirow{2}{*}{$448\times 448\times 3$} & 3 seconds & $67.62\%$\\\cline{3-4}
 &  & 10 seconds & $75.37\%$\\\hline
\multirow{6}{*}{CosFace} & \multirow{2}{*}{$224\times 224\times 3$} & 3 seconds & $66,51\%$\\\cline{3-4}
 &  & 10 seconds & $79,42\%$\\\cline{2-4}
 & \multirow{2}{*}{$432\times 288\times 3$} & 3 seconds & $69,82\%$\\\cline{3-4}
 &  & 10 seconds & $\pmb{83,15\%}$\\\cline{2-4}
 & \multirow{2}{*}{$448\times 448\times 3$} & 3 seconds & $64,93\%$\\\cline{3-4}
 & & 10 seconds & $79,56\%$\\\hline
\multirow{6}{*}{ArcFace} & \multirow{2}{*}{$224\times 224\times 3$} & 3 seconds & $66,28\%$\\\cline{3-4}
 &  & 10 seconds & \pmb{$81,33\%$}\\\cline{2-4}
 & \multirow{2}{*}{$432\times 288\times 3$} & 3 seconds & $68,05\%$\\\cline{3-4}
 &  & 10 seconds & $80,79\%$\\\cline{2-4}
 & \multirow{2}{*}{$448\times 448\times 3$} & 3 seconds & $64,15\%$\\\cline{3-4}
 &  & 10 seconds & $79,32\%$\\
\hline
\end{tabular}}
\label{resultstab}
\end{table}

\subsection{Effect of Mel-Spectrogram dimensions} \label{Effect of dim}

The experimental results also underscore the significant influence of mel-spectrogram dimensions on the model's top-1 accuracy in SI. Among the various configurations evaluated, the mel-spectrogram dimensions of $432\times 288\times 3$ almost always yielded the highest top-1 accuracy, achieving a peak performance of $83.15\%$. Following closely, the dimensions of $224\times 224\times 3$ secured the second-highest accuracy, while the $448\times 448\times 3$ dimensions were ranked third.

This discernible performance variation across different input sizes suggests that not only the overall resolution but also the aspect ratio of the mel-spectrograms plays a critical role in capturing the spectral and temporal characteristics that distinguish speakers. It is hypothesized that the $432\times 288\times 3$ configuration may provide a more effective balance, allowing the Convolutional Neural Network (CNN) to better learn the intricate patterns within the speech signal that are indicative of speaker identity. This finding emphasizes that simply resizing mel-spectrograms to larger dimensions does not guarantee improved performance; rather, optimizing the dimensions to suit the underlying data structure is crucial for boosting accuracy. Consequently, selecting appropriate mel-spectrogram dimensions is a key factor in maximizing the effectiveness of our modified VGG16 model for SI. 

\subsection{Impact of loss functions} \label{Impact of LF}

CosFace demonstrated clear superiority over both ArcFace and traditional Softmax across most configurations. 
CosFace achieved the highest identification accuracy of $83.15\%$, representing a $2.36$ percentage point improvement over ArcFace's best performance ($80.79\%$) and a 6.74 percentage point advantage over traditional Softmax ($76.41\%$) when using optimal conditions.

The superior performance of CosFace can be attributed to its ability to enforce a fixed angular margin combined with a scaling factor, which creates more discriminative decision boundaries and pushes speaker embeddings further apart in the feature space.

Unlike traditional Softmax that fails to explicitly optimize for class separation, CosFace addresses fundamental limitations by directly enhancing intra-class compactness while increasing inter-class diversity.

While ArcFace also incorporates margin-based constraints through geodesic distance optimization, it demonstrated more variability across configurations. Notably, ArcFace achieved its peak performance of $81.33\%$ with $244\times 244\times 3$ dimensions and ten-second clips, but consistently underperformed compared to CosFace when using the optimal $432\times 288\times 3$ configuration.

\section{Conclusion} \label{Conclusion}

Our study comprehensively achieved all stated objectives. 
\begin{itemize}
    \item We successfully evaluated the comparative effectiveness of the ArcFace and CosFace loss functions against traditional Softmax loss and identified their superior performance when used with our modified VGG16 architecture.
    \item We developed and validated an audio looping technique that effectively extends shorter audio samples, proving that temporal information enhancement can overcome duration limitations in practical applications. 
    \item We systematically identified optimal mel-spectrogram dimensions (432×288×3) that balance computational efficiency with identification accuracy.
\end{itemize}

\subsection{Practical Implications and Applications} \label{practical}

The enhanced accuracy and robustness achieved through our approach could benefit security systems requiring reliable voice authentication, improve virtual assistant personalization capabilities, and support forensic voice analysis applications. The audio looping technique particularly addresses practical challenges where only short utterances are available, making the system more applicable to real-world scenarios where extended speech samples are not always obtainable.

\subsection{Study Limitations and Considerations}

It is important to contextualize objectives and findings of the present study within its intended research scope. This investigation was not designed as an attempt to achieve state-of-the-art performance on the VoxCeleb1 dataset, but rather as a systematic exploration of audio looping behavior across different system configurations. Our primary research objective centered on understanding how temporal augmentation through audio looping interacts with various loss functions and input dimensions, providing foundational insights for future optimization efforts.

The experimental design prioritized comprehensive configuration space exploration over performance maximization. By systematically evaluating audio looping across three distinct loss functions (Softmax, ArcFace, CosFace), multiple mel-spectrogram dimensions ($224\times 224\times 3$, $448\times 448\times 3$, $432\times 288\times 3$), and two temporal durations (3-second, 10-second), we established a controlled framework for understanding the behavioral characteristics of our proposed augmentation method. This systematic approach enables reliable conclusions about the relative effectiveness of different configurations and provides guidance for practitioners considering similar approaches.

The choice to maintain consistent architectural and preprocessing approaches across all experiments, while potentially limiting absolute performance, ensures that observed differences can be attributed to the specific variables under investigation rather than confounding factors.

Our comparison with existing literature serves as a contextual framework for understanding where audio looping fits within the broader spectrum of augmentation strategies. The competitive performance achieved relative to some established methods (notably matching Yadav and Rai's 83.5\% within 0.35\%\cite{yadav2018learning}) demonstrates that audio looping represents a viable augmentation approach worthy of further development.

The identification of optimal input dimensions ($432\times 288\times 3$) and the superior performance of CosFace over ArcFace in this context represent practical contributions that can inform future system design decisions, even as absolute performance optimization remains a goal for subsequent research phases.

\subsection{Future Research Directions}

The audio looping technique warrants systematic investigation to establish optimal operational parameters. Questions include determining the minimum viable audio duration for effective looping and the maximum extension length before performance plateaus. Future research should explore alternative looping strategies beyond simple repetition, such as partial overlap looping or intelligent looping that prioritizes high-energy audio segments. Additionally, adaptive algorithms that adjust looping patterns based on spectral content analysis could yield superior results compared to full-segment repetition.

Combining audio looping with established techniques could address multiple system limitations simultaneously. Research should investigate optimal combinations of audio looping with random sampling for class balancing, noise injection for robustness, and other modifications. Adaptive systems that select augmentation strategies based on per-speaker data availability represent a particularly promising direction for addressing both class imbalance and temporal constraints within unified frameworks.

The findings regarding network depth suggest systematic architectural investigation could yield improvements. Future research should explore modern architectures—including transformers and efficient convolutional designs—specifically combined with audio looping. Additionally, developing networks explicitly designed to leverage the temporal patterns created by looping could achieve superior performance compared to standard architectures applied to extended inputs.

Systematic evaluation across datasets featuring different languages, acoustic conditions, and speaker demographics would establish the method's robustness and identify potential limitations. Cross-linguistic studies examining effectiveness across different phonetic structures and prosodic patterns are particularly important for understanding universal applicability.

Extension to open-set SI scenarios would broaden practical applicability, though this requires investigating how temporal extension affects unknown speaker rejection capabilities. The development of threshold optimization strategies and uncertainty quantification techniques specifically designed for looped audio inputs could address deployment challenges while maintaining identification benefits.

\subsection{Concluding Remarks}

In conclusion, this research provides a systematic evaluation of margin-based loss functions in conjunction with an audio looping data augmentation strategy, within the framework of CNN-based SI. Achieving a top-1 accuracy of 83.15\% on the VoxCeleb1 dataset, our system delivers competitive performance within the existing literature range. The primary value of this work lies in the proposed audio looping technique which, despite its simplicity, offers a practical and easily implementable solution for extending limited audio samples. This method is particularly well-suited for resource-constrained applications, and its integration with advanced loss functions demonstrates a viable pathway for improving SI accuracy in scenarios with short-duration recordings.

\bibliographystyle{splncs04}
\bibliography{citations}

\begin{thebibliography}{10}
\providecommand{\url}[1]{\texttt{#1}}
\providecommand{\urlprefix}{URL }
\providecommand{\doi}[1]{https://doi.org/#1}

\bibitem{9519395}
Abdullah, H., Warren, K., Bindschaedler, V., Papernot, N., Traynor, P.: Sok: The faults in our asrs: An overview of attacks against automatic speech recognition and speaker identification systems. In: 2021 IEEE Symposium on Security and Privacy (SP). pp. 730--747 (2021). \doi{10.1109/SP40001.2021.00014}

\bibitem{an2019deep}
An, N.N., Thanh, N.Q., Liu, Y.: Deep cnns with self-attention for speaker identification. IEEE access  \textbf{7},  85327--85337 (2019)

\bibitem{anand2019shotspeakerrecognitionusing}
Anand, P., Singh, A.K., Srivastava, S., Lall, B.: Few shot speaker recognition using deep neural networks (2019), \url{https://arxiv.org/abs/1904.08775}

\bibitem{badshah2019deep}
Badshah, A.M., Rahim, N., Ullah, N., Ahmad, J., Muhammad, K., Lee, M.Y., Kwon, S., Baik, S.W.: Deep features-based speech emotion recognition for smart affective services. Multimedia Tools and Applications  \textbf{78},  5571--5589 (2019)

\bibitem{campbell2002speaker}
Campbell, J.P.: Speaker recognition: A tutorial. Proceedings of the IEEE  \textbf{85}(9),  1437--1462 (2002)

\bibitem{deng2019arcface}
Deng, J., Guo, J., Xue, N., Zafeiriou, S.: Arcface: Additive angular margin loss for deep face recognition. In: Proceedings of the IEEE/CVF conference on computer vision and pattern recognition. pp. 4690--4699 (2019)

\bibitem{furui2010speaker}
Furui, S.: Speaker recognition in smart environments. In: Human-centric interfaces for ambient intelligence, pp. 163--184. Elsevier (2010)

\bibitem{guo2021speaker}
Guo, M., Yang, J., Gao, S.: Speaker recognition method for short utterance. In: Journal of physics: conference series. vol.~1827, p. 012158. IOP Publishing (2021)

\bibitem{hajibabaei2018unified}
Hajibabaei, M., Dai, D.: Unified hypersphere embedding for speaker recognition. arXiv preprint arXiv:1807.08312  (2018)

\bibitem{juang2003digital}
Juang, B.H., Sondhi, M.M., Rabiner, L.R.: Digital speech processing  (2003)

\bibitem{kamruzzaman2010speaker}
Kamruzzaman, S., Karim, A., Islam, M.S., Haque, M.E.: Speaker identification using mfcc-domain support vector machine. arXiv preprint arXiv:1009.4972  (2010)

\bibitem{lambamo2022analyzing}
Lambamo, W., Srinivasagan, R., Jifara, W.: Analyzing noise robustness of cochleogram and mel spectrogram features in deep learning based speaker recognition. applied sciences  \textbf{13}(1), ~569 (2022)

\bibitem{lin2022cross}
Lin, Y., Qin, X., Li, M.: Cross-domain arcface: Learnging robust speaker representation under the far-field speaker verification. In: Proc. FFSVC 2022. pp.~6--9 (2022)

\bibitem{lu2008investigation}
Lu, X., Dang, J.: An investigation of dependencies between frequency components and speaker characteristics for text-independent speaker identification. Speech communication  \textbf{50}(4),  312--322 (2008)

\bibitem{magalhaes2021voice}
Magalh{\~a}es, A.F.d.S., et~al.: Voice recognition of users for virtual assistant in industrial environments. Master's thesis (2021)

\bibitem{morandieffect}
MORANDI, D.A.: Effect of pitch modification on the voice identification of the speakers

\bibitem{nagrani2017voxceleb}
Nagrani, A., Chung, J.S., Zisserman, A.: Voxceleb: a large-scale speaker identification dataset. arXiv preprint arXiv:1706.08612  (2017)

\bibitem{nugroho2022enhanced}
Nugroho, K., Noersasongko, E., et~al.: Enhanced indonesian ethnic speaker recognition using data augmentation deep neural network. Journal of King Saud University-Computer and Information Sciences  \textbf{34}(7),  4375--4384 (2022)

\bibitem{qawaqneh2017deep}
Qawaqneh, Z., Mallouh, A.A., Barkana, B.D.: Deep neural network framework and transformed mfccs for speaker's age and gender classification. Knowledge-Based Systems  \textbf{115},  5--14 (2017)

\bibitem{saritha2024deep}
Saritha, B., Laskar, M.A., Kirupakaran, A.M., Laskar, R.H., Choudhury, M., Shome, N.: Deep learning-based end-to-end speaker identification using time--frequency representation of speech signal. Circuits, Systems, and Signal Processing  \textbf{43}(3),  1839--1861 (2024)

\bibitem{sharif2025improving}
Sharif-Noughabi, M., Razavi, S., Mohamadzadeh, S.: Improving the performance of speaker recognition system using optimized vgg convolutional neural network and data augmentation. International Journal of Engineering  \textbf{38}(10),  2414--2425 (2025)

\bibitem{simonyan2014vgg}
Simonyan, K., Zisserman, A.: Very deep convolutional networks for large-scale image recognition. arXiv preprint arXiv:1409.1556  (2014)

\bibitem{sinith2010novel}
Sinith, M., Salim, A., Sankar, K.G., Narayanan, K.S., Soman, V.: A novel method for text-independent speaker identification using mfcc and gmm. In: 2010 International Conference on Audio, Language and Image Processing. pp. 292--296. IEEE (2010)

\bibitem{tang2022bimodal}
Tang, Y., Hu, Y., He, L., Huang, H.: A bimodal network based on audio--text-interactional-attention with arcface loss for speech emotion recognition. Speech Communication  \textbf{143},  21--32 (2022)

\bibitem{wang2018additive}
Wang, F., Cheng, J., Liu, W., Liu, H.: Additive margin softmax for face verification. IEEE Signal Processing Letters  \textbf{25}(7),  926--930 (2018)

\bibitem{wang2018cosface}
Wang, H., Wang, Y., Zhou, Z., Ji, X., Gong, D., Zhou, J., Li, Z., Liu, W.: Cosface: Large margin cosine loss for deep face recognition. In: Proceedings of the IEEE conference on computer vision and pattern recognition. pp. 5265--5274 (2018)

\bibitem{yadav2018learning}
Yadav, S., Rai, A.: Learning discriminative features for speaker identification and verification. In: Interspeech. pp. 2237--2241 (2018)

\bibitem{zouhir2025power}
Zouhir, Y., Zarka, M., Ouni, K., El~Amraoui, L.: Power wavelet cepstral coefficients (pwcc): An accurate auditory model-based feature extraction method for robust speaker recognition. IEEE Access  (2025)

\end{thebibliography}

\end{document}